\documentclass[prc,aps,showpacs,floatfix,nofootinbib,twocolumn,letterpaper]{revtex4}
\usepackage{graphicx}
\usepackage[T1]{fontenc}
\usepackage[latin9]{inputenc}
\usepackage{textcomp}
\usepackage{amsmath}
\usepackage{amssymb}
\def\be{\begin{equation}}
\def\ee{\end{equation}}

\def\u{$^{236}$U}
\def\eps{\varepsilon}
\def\gpre{$|166\rangle$}
\def\gpost{$|168p\rangle$}
\def\neck{$N_{\rm neck}$}
\begin{document}

\title{Diabatic scission paths}

\author{G.F.~Bertsch}
\affiliation{Institute for Nuclear Theory and Department of Physics,
University of Washington, Seattle, Washington, USA}
\email{bertsch@uw.edu}
\author{W. Younes}
\affiliation{ 
Lawrence Livermore National Laboratory, Livermore, CA 94551, USA }
\author{L.M.~Robledo}
\affiliation{Departamento de Fisica Teorica, Universidad Aut\'onoma de Madrid, E-28049
Madrid, Spain}

\begin{abstract}
An outstanding problem in the theory of nuclear fission is to understand
the Hamiltonian dynamics at the scission point.  
In this work the fissioning nucleus is modeled in self-consistent
mean-field theory as a set of Generator Coordinate (GCM) configurations
passing through the scission point.  In contrast to previous methods,
the configurations are constructed in the Hartree-Fock approximation
with axially symmetric mean fields and using the 
$K$-partition numbers as additional constraints.  The goal of this
work is to find paths through the scission point where the overlaps
between neighboring configurations are large.  A measure of distance along the 
path is proposed that is insensitive to the division of the path into
short segments.
For most of the tested $K$-partitions two shape degrees of freedom are
adequate to define smooth paths.  However, some of the configurations and
candidate paths have sticking points where there are substantial
changes in the many-body 
wave function, especially if quasiparticle excitations are present. 
The excitation energy deposited in fission fragments arising from
thermal excitations in the pre-scission configurations is determined
by tracking orbital occupation numbers along the scission paths.
This allows us to assess the validity of the well-known
scission-point statistical model, in which the scission process is
assumed to be fully equilibrated up to the separated fission fragments.
The nucleus \u~is taken as a representative example in the calculations.
\end{abstract}
 
\maketitle

\section{Introduction}

The final step of nuclear fission, namely the scission into two 
(or more) distinct fragments, has always been difficult to
understand in fully microscopic models. In particular, self-consistent
mean-field theory has been very successful in treatment many aspects
of fission theory but has shed little light on the final scission
dynamics.  The problem can be seen in typical constructions of
fission paths by the Generator Coordinate Method (GCM) \cite{be03}.  This
involves a constrained minimization of
the mean-field configurations, treating the expectation value 
of each constraining field as 
a coordinate.  Most important among the coordinates is the
elongation of the system.  Configurations along the 
fission path are step-wise defined by re-minimizing the previous 
configuration at a  slightly larger elongation, but more sophisticated
propagation methods are possible \cite{sc16,re16,le18}.  At the scission point 
these procedures break down:  the re-minimization produces
a configuration very different from the previous one.
Efforts to construct a continuous path
have often focused on introducing more shape constraints,
but the difficulties remain \cite{du12}.  

In this work we will also follow the GCM approach, but with some differences
from previous work.  We will use the Hartree-Fock rather 
Hartree-Fock-Bogoliubov approximation to represent the 
configurations.  We assume that the mean-field
potential is axially symmetric, so the angular momentum of the
orbitals about the fission axis $K$ is conserved.  
Then the
dynamics conserves the number of particles in
orbitals of a given $K$. The set consisting of number of occupied orbits 
for $K$ will be called the {\it $K$-partition}.  
The resulting dynamics, preventing  particles from jumping orbitals, is
called\footnote{ See Refs. \cite{ya12,ta16,sc18}
for examples of its application in other research fields.} {\it diabatic}
\cite{na93,sc81}.
In Ref. \cite{be18} we have
explored some of the diabatic configurations in \u~ leading to scission.
In this work we study in more detail
the changes in the wave functions and energies going through
the scission point.

There are several characteristics of a diabatic path through scission
that we will examine.  The first is to determine the difficulty in
defining a smooth path in terms of the number of shape constraints
need.  Our original hope was that a single elongation constraint
would suffice, but that turns out not to be adequate.  We also define
and evaluation a measure of the length of the path through the
scission point; long path are more difficult to traverse and the
quantum theory would require more GCM configurations to describe
them.    An important physical question
is how the scission dynamics affects the energies in the
final state fragments.  The configurations on the fission
path  
are the lowest energy ones subject to the constraints,
which we call the zero-quasiparticle configurations (ZQP).
The excitation energy above the ZQP configuration can be affected
by the scission in two ways.  First, the diabatic dynamics could
induce quasiparticle excitations, thereby increasing the total
excitation energy in the final state.  Even if this
does not occur, the diabatic evolution will change the excitation
energy because the quasiparticle energies will change.  Also,
the excitation energy sharing between fragments is determined
at the scission point.
In Sect. II below we take the example of the configuration 
named Glider in Ref. \cite{be18} to explain how we calculate
the various properties of interest.  
An important question is how much of the behavior seen for Glider
is generic with respect to different $K$-partitions or different
energy functionals.  In Sect. III we analyze several more 
$K$-partitions with two quite different energy functionals
to see what general conclusions can be made.  One particular
question is how well statistical approaches to the
scission dynamics can be justified in a microscopic approach.
There has been considerable success of the
scission-point statistical
model \cite{wi76,le15,al18} which assumes that the excitation energy is
fully
equilibrated between the two fragments at some fixed separation
between their surfaces. In particular, Ref. \cite{al18} finds good
agreement with experiments sensitive to the energy sharing.

\section{Glider}

We first summarize how the GCM scission configurations were constructed
in Ref. \cite{be18}.  The calculations are carried 
out\footnote{See Supplementary Materials \cite{supp} for sample wave functions and
the codes used to analyze them.}
with the
code HFBaxial \cite{LMR} using the Gogny D1S energy functional \cite{be91}.  
The code finds minima in the
Hartree-Fock-Bogoliubov energy functional constrained by the expectation
values of external fields that serve as generator coordinates. 
The GCM fields available
are the mass multipole moments\footnote{Note that $Q_2$ defined here is one-half the conventional
definition.}
\be
\hat Q_L = r^L P_L(\cos \theta),
\ee 
and fields associated with particle number and its fluctuations in 
the HFB wave functions.
The code assumes that the single-particle Hamiltonian is axially 
symmetric.  
To find typical scission configurations, we constructed   
a 
fission path by  HFB minimizations with only one shape
constraint, 
namely the mass quadrupole moment $Q_2$.  The minimizations were carried
out iteratively starting from the ground-state configuration. 
At each cycle in the iteration, the $Q_2$ was increased by a
small amount using the previous minimum as the starting configuration.
For the nucleus \u, the configuration underwent a major rearrangement
at $Q_2\approx 168$ b.  At that point the shape changed abruptly
with a near disappearance of the neck joining the two proto-fragments.
To get a closer view of the wave function dynamics at that point
we determined the dominant Hartree-Fock configuration in the 
HFB wave functions. This was carried out by re-minimization
of the HFB configuration with an added constraint on the number operator
$\hat N^2$,  effectively turning off the pairing field.  Since the 
single-particle Hamiltonian is axially symmetric, the HF configurations
can be characterized by its $K$-partition as well as by its density
moments.  

Among the HF states found, Glider is an intermediate configuration just
at the edge of the scission point; its $K$-partition is given in Table I.
\begin{table}[htb]  
\begin{center}  
\begin{tabular}{|c|cccccc|cccccc|}  
\hline  
$K$-partition                &        \multicolumn{6}{c|}{protons}       & 
\multicolumn{6}{c|}{neutrons } \\
\hline  
$2K$  &   1 & 3 & 5 & 7 & 9 & 11  & 1 & 3 & 5 & 7 & 9 & 11 \\  
\hline  
Glider     & 22 & 14 & 6 &3 &1 & 0& 31 & 20 & 11 & 6 &3 &1\\ 
A      & 22 & 14 & 7 &3  &0 & 0& 31 & 21 & 12 & 6 &2 &0 \\ 
B      & 21 & 13 & 7 &4  &1 & 0& 30 & 20 & 11 & 7 &3 &1 \\ 
C      & 21 & 14 & 7 &3 &1 & 0& 30 & 21 & 11 & 6 &3 &1 \\ 
\hline  
\end{tabular}  
\caption{$K$-partition of the configuration Glider and others to
be discussed in Sec. \ref{other}.  The entries
are the number of occupied orbits of a given $K>0$.  The total
number of particles of given $|K|$ is twice that.
\label{NK}  
}
\end{center}  
\end{table}   
From
that point the HFB fission path jumps abruptly to a configuration of separated
fragments with very different shapes.  The HF reduction shows that
this is accompanied by a major rearrangement of orbital occupancies.
However, constraining the $K$-partition allows
one to track Glider over a wide range of deformations going into
post-scission shapes.  Its energy as a function of deformation is shown
in Fig. 1.  One can see a transition at $Q_2= 168-170$ b where the
neck disappears. Interestingly, there are two local minima at $Q_2=168$
and 170 b in the HF energy surface when constrained only by the $Q_2$ moments.  The higher energy
configurations are obtained by stepping from smaller to larger $Q_2$
values, and the lower ones by stepping in the opposite direction.
Beyond these two $Q_2$ points, the iteration gives
identical configurations in both directions.
\begin{figure}[tb] 
\begin{center} 
\includegraphics[width=1.0\columnwidth]{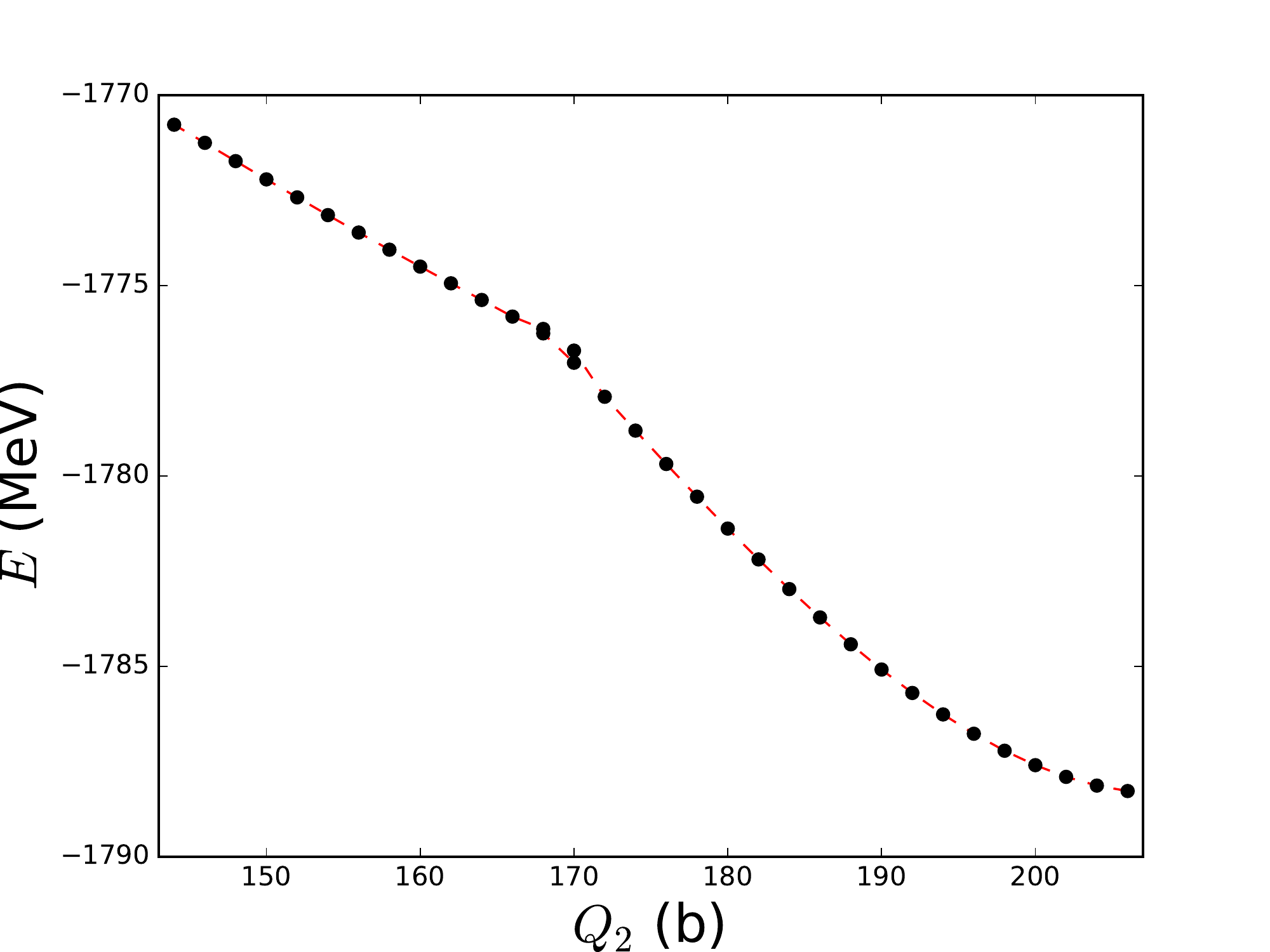} 
\caption{
\label{QE} 
Energy of Glider with the D1S energy functional and constrained by the
quadrupole field $Q_2$.
}
\end{center}
\end{figure} 
The density distributions of the $Q_2=166$ b configuration $|166\rangle$ and the 
post-scission $Q_2=168$ b configuration $|168p\rangle$ are shown in Fig. 2. 
These
configurations have a substantial overlap  ($\langle 166|168p\rangle
= 0.29$) and there is no 
obvious change in structure
besides the disappearance of the neck.
\begin{figure}[tb] 
\begin{center} 
\includegraphics[width=1.0\columnwidth]{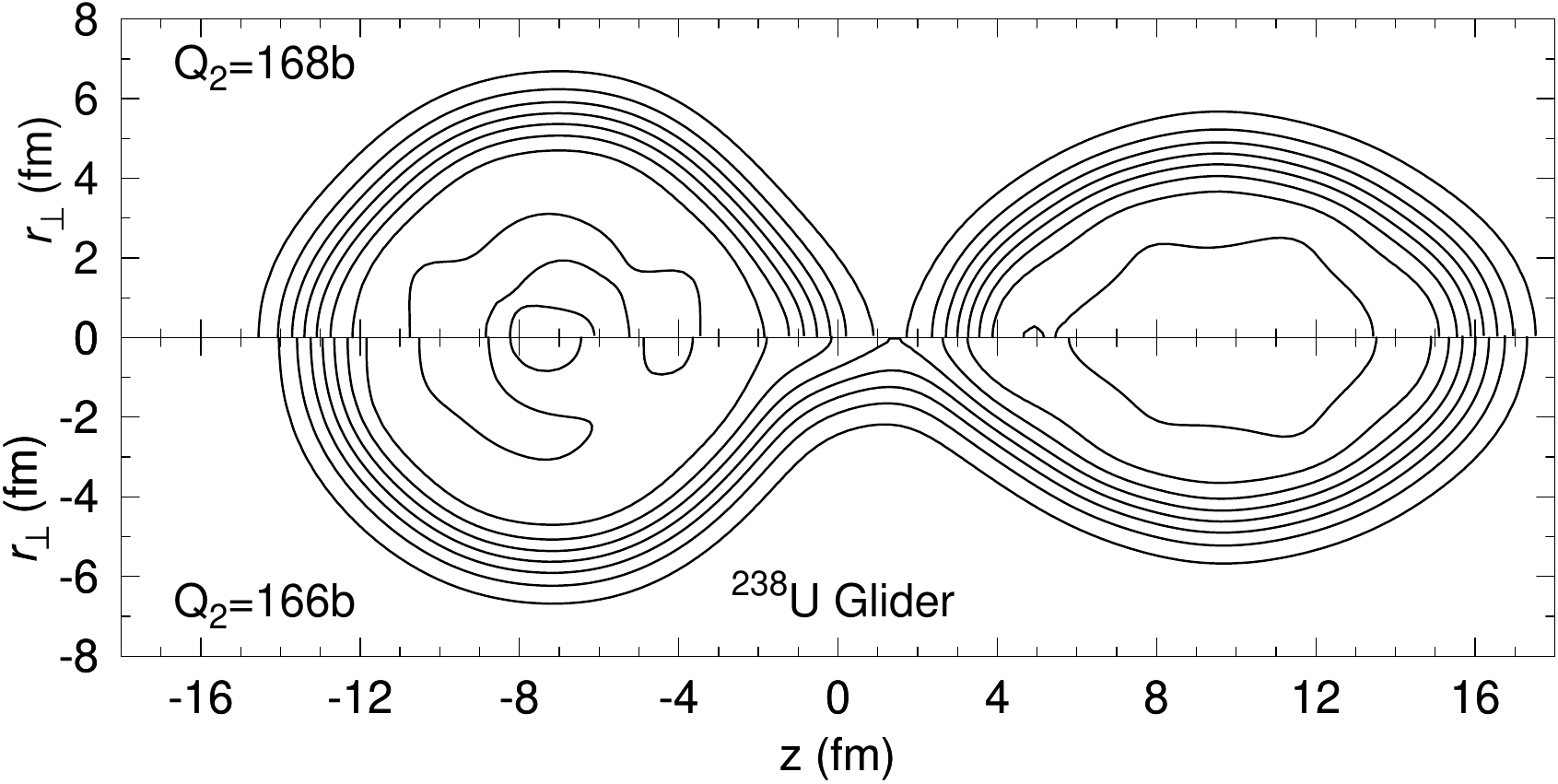} 
\caption{
\label{rho-plot} 
Density distributions of the Glider configurations at  two
HF minima at the scission point.  Top panel:  configuration $|166\rangle$; bottom panel:
$|168p\rangle$.
Contour lines of the mass density are spaced 
by $\Delta \rho = 0.016$ fm$^{-3}$.}
\end{center}
\end{figure} 
To analyze the transition in more detail, we will introduce an additional
GCM coordinate based on a field sensitive to the number of nucleons 
in the neck region\cite{wa02}.  The definition is 
\be
\hat N_{\rm neck} = \exp\left(-(z-z_A)^2/a_N^2\right),
\ee
with $z_A$ chosen at the minimum point along the fission axis $z$ and
$a_N = 1$ fm.  Its expectation value as function of $Q_2$
is shown in Fig. \ref{n_neck}. One sees that there is a discontinuous change
in $N_{\rm neck}$ from $\sim 1.7$ to $\sim 0.6$ where there are two local
minima. 
\begin{figure}[tb] 
\begin{center} 
\includegraphics[width=1.0\columnwidth]{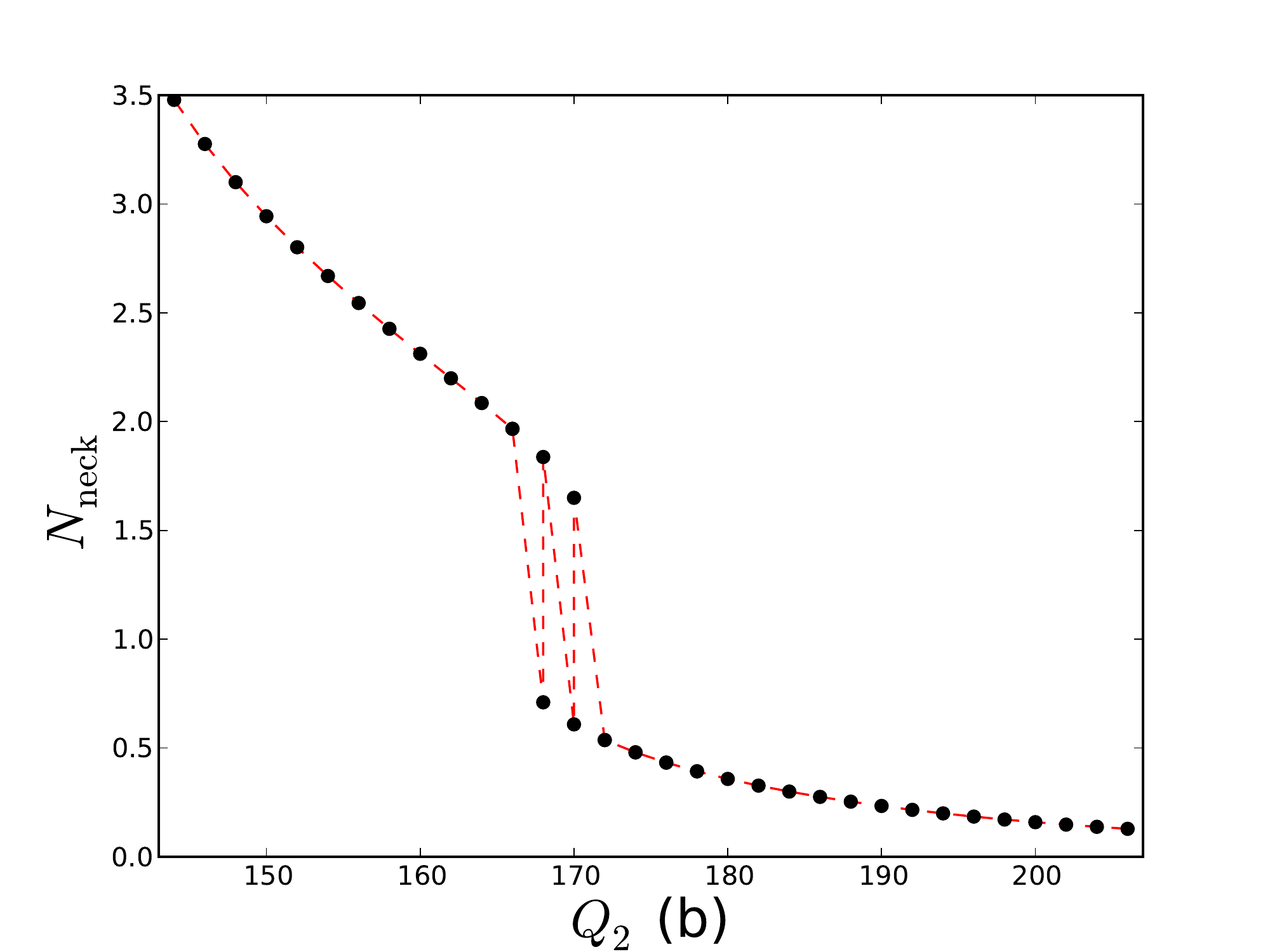} 
\caption{
\label{n_neck}
Neck  parameter $N_{\rm neck}$ as a function of deformation for
Glider.  The circles starting from the upper left are the HF
minima constrained by $Q_2$ stepping from the previous
solution at $Q_2- 2$ b.  The minima starting from the lower
right are similarly constrained stepping from the previous solution
at $Q_2+2$ b.  There are two distinct minima at $Q_2=168$ and
170 b. 
}
\end{center}
\end{figure} 
In Table II we show some of the characteristics
of the two solutions.  
\begin{table}[htb]  
\begin{center}  
\begin{tabular}{|c|cc||cc|}  
\hline  
             &   $|166\rangle$   & $|168p\rangle$   & heavy & light \\
\hline  
$Z$  & 92 & 92 &  52 & 40\\
$N$  & 144 & 144 & 84 & 60 \\
$E$ (MeV) &-1775.8 & -1776.1   & & \\
$Q_2$ (b) & 166   & 168    & 1.5 & 4.8 \\
$Q_3$ (b$^{3/2})$& 50.2  & 51.4  & 0.06 & 0.06 \\
$N_{\rm neck}$ &  2.0 & 0.7   && \\
\hline
\end{tabular} 
\caption{Properties of the configuration Glider at the pre-scission
point $Q_2 = 166$, and the post-scission solution  
at $Q_2 = 168$ b.  The last two columns report properties of the
daughter nuclei, extracted from the density distribution of 
the post-scission configuration at 168 b. 
\label{glider}  
}
\end{center}  
\end{table}   
A continuous scission path can be constructed by adding $\hat N_{\rm neck}$ as a generator
coordinate.  For example, we can define a unique intermediate configuration
$|167m\rangle$ 
by constraining  $Q_2 = 167$ b and $N_{\rm neck} =  1.1$, as both
$|166\rangle$ and
$|168p\rangle$ converge to it when the wave functions are re-minimized
with the new constraint.  The overlaps of the configurations are given in 
Table \ref{overlaps}.
\begin{table}[htb]  
\begin{center}  
\begin{tabular}{|c|ccc|}  
\hline  
 Overlaps     &   166  & 167m   & 168p\\
\hline  
166 & 1.0 & 0.79 &  0.29 \\
167m  & 0.79 & 1.0 & 0.67 \\
168p & 0.29 & 0.67   & 1.0 \\
\hline
\end{tabular} 
\caption{Overlaps of Glider configurations near the scission point.
\label{overlaps}  
}
\end{center}  
\end{table}   

It will be useful to have a cumulative measure of the overlaps along
the scission path that is insensitive to the details of the step
sizes used to construct the path.  This is achieved by the distance
function $\zeta$ defined as
\be
\label{zeta}
\zeta =\sum_{n=1}^{N-1} \left(-\log |\langle n | n+1\rangle|\right)^{1/2}. 
\ee
for diabatic paths consisting of a chain of $N$ configurations $|n\rangle$.
 we
Applying Eq. \ref{zeta} to the path between end configurations in 
Table \ref{overlaps}, we find
$\zeta= 1.13,1.12$, and 1.13 with 0,1, and 3 intermediate configurations.
Clearly this satisfies our insensitivity demand.

\subsection{Orbitals}
Here we examine properties of the HF orbitals and how they evolve during
the scission.  An important goal is to 
calculate the changes in excitation energy associated with the scission and how
that energy is distributed between the final state fragments.  Up to now we have only treated ZQP
configurations, but the theory can be easily extend by 
allowing partial occupation numbers $n_\alpha$ for orbitals $\alpha$
in the vicinity of the Fermi level.  In the independent-quasiparticle 
approximation the excitation energy $E^*$ above the ZQP value is given by
\be  
E^* = \sum_\alpha \eps_\alpha \left(n_\alpha - n_\alpha^0\right)
\label{Estar-eq}
\ee
where $n_\alpha^0=0$ or 1 is the occupation number in the ZQP
configuration.  In the diabatic approximation, 
the orbital energies will change but the
occupation factors will be frozen at the their initial values.

\begin{figure}[tb] 
\begin{center} 
\includegraphics[width=1.0\columnwidth]{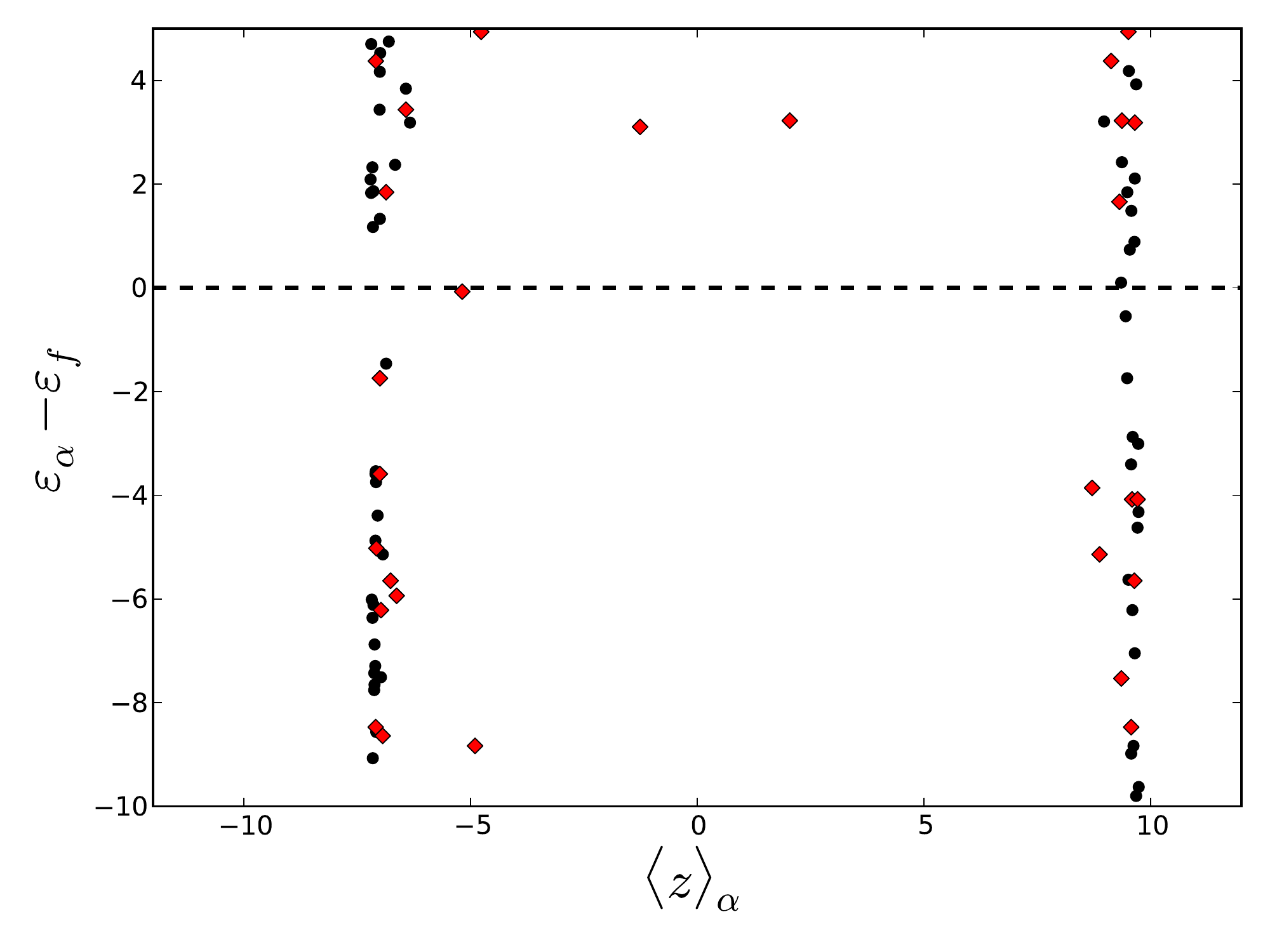} 
\includegraphics[width=1.0\columnwidth]{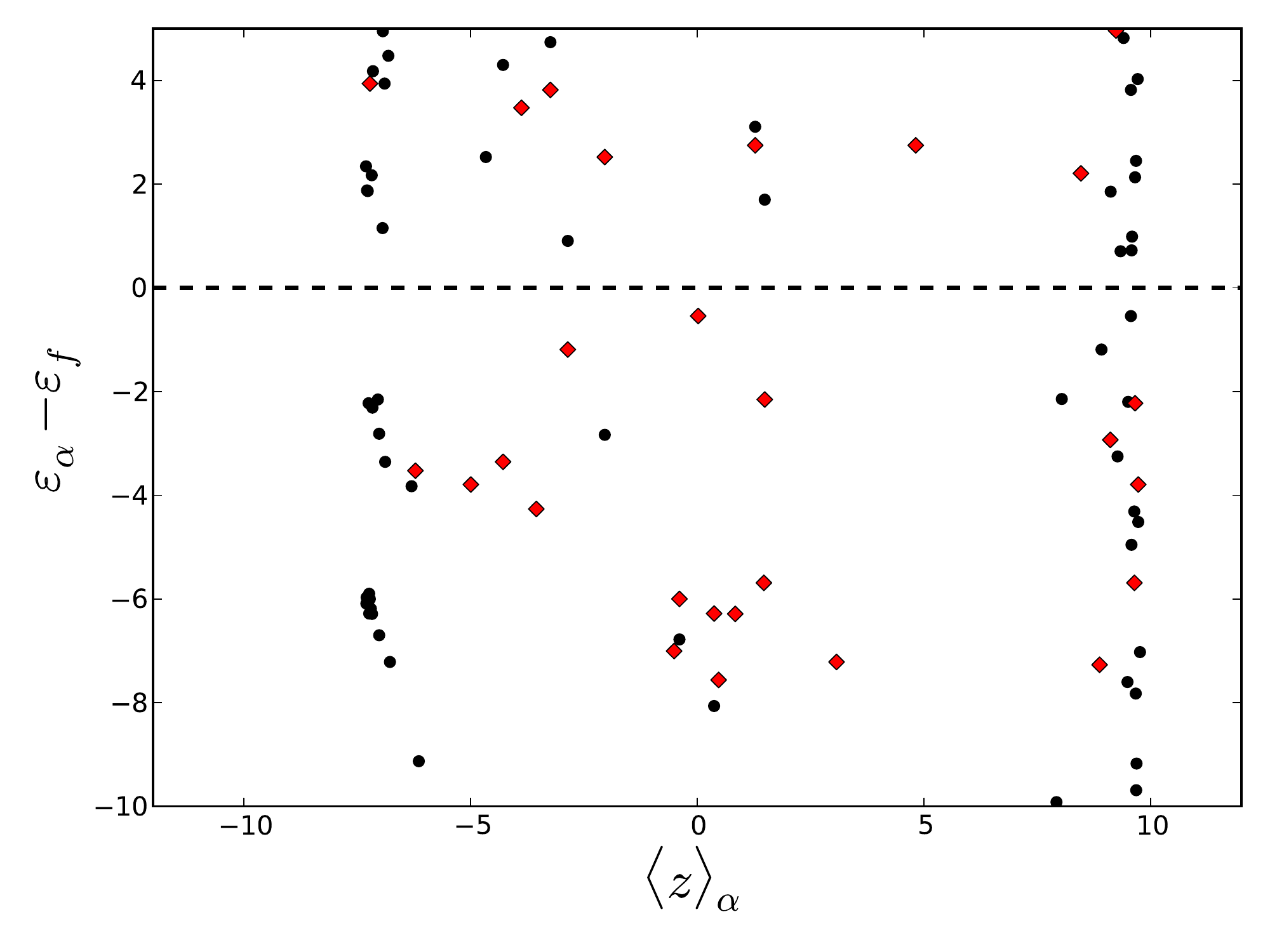} 
\caption{Orbital locations and their energies 
as a function of $\langle z \rangle$.  Upper panel:  $|168p\rangle$;
 lower panel: $|166\rangle$.
Orbitals with $K=1/2$ are shown as red diamonds; others by black circles.
respectively. Energies are with respect to the Fermi energy $\eps_f$.
Units are MeV for energies and fm for $\langle Z\rangle$.
\label{eps-vs-z} 
}
\end{center} 
\end{figure}

To calculate the sharing of excitation energy between the two
fission fragments,  
we need to understand the
localization of the orbitals onto one fragment or the other.
A rough indicator the expectation value
of the orbital density along the fission axis, 
\be
\langle z\rangle_\alpha = \int d^3 r\, z |\phi_\alpha(\vec r)|^2
\ee
Fig. \ref{eps-vs-z} shows the orbital energies and
their $z$-averaged positions for the configurations $|166\rangle$ and
$|168p\rangle$, covering the energy band $-12 <  \varepsilon -\varepsilon_f
< 5$ Mev.
The more deeply bound orbitals have $\langle z\rangle$ 
close to $-7.3$ or $9.6$ fm, corresponding to
the cm positions of the heavy or light fragment, respectively.  For the
post-scission configuration 
(upper panel of Fig. \ref{eps-vs-z}), practically all of 
the occupied orbitals follow that pattern.   The situation is quite
different for the pre-scission configuration shown in the lower panel. 
There are 10-20 orbitals that have much smaller $\langle z\rangle$,
indicating a significant probability on both proto-fragments.  One would
expect that the extent of the bridging between the two fragments would
depend strongly on $K$: orbitals with high $K$ quantum numbers have small
densities near the fission axis and would not have a substantial presence in
the neck region.  This is confirmed by the data shown in the Figure.  One
can see that most of the bridge orbitals have $K=1/2$.

\begin{figure}[tb] 
\begin{center} 
\includegraphics[width=1.0\columnwidth]{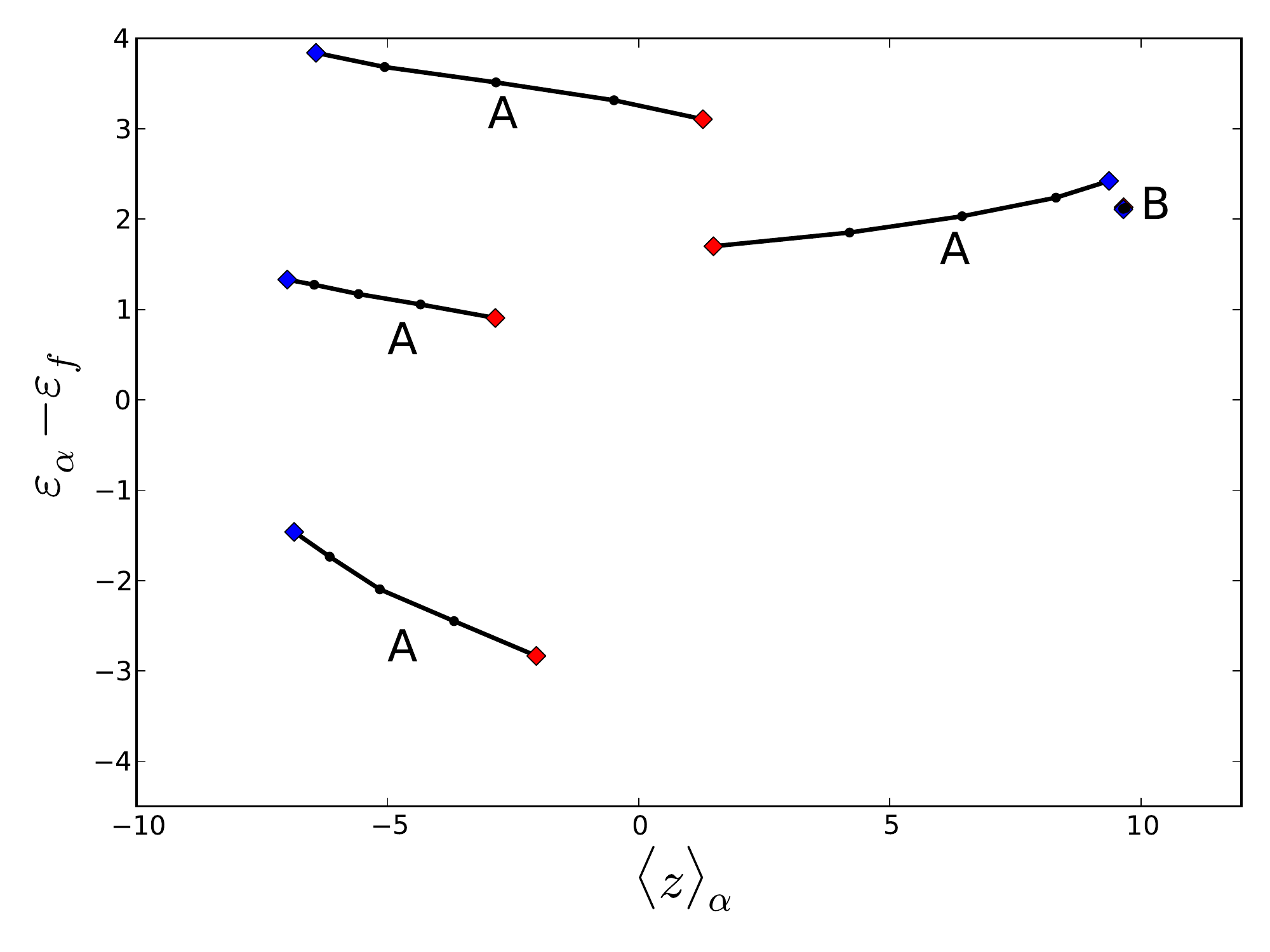} 
\includegraphics[width=1.0\columnwidth]{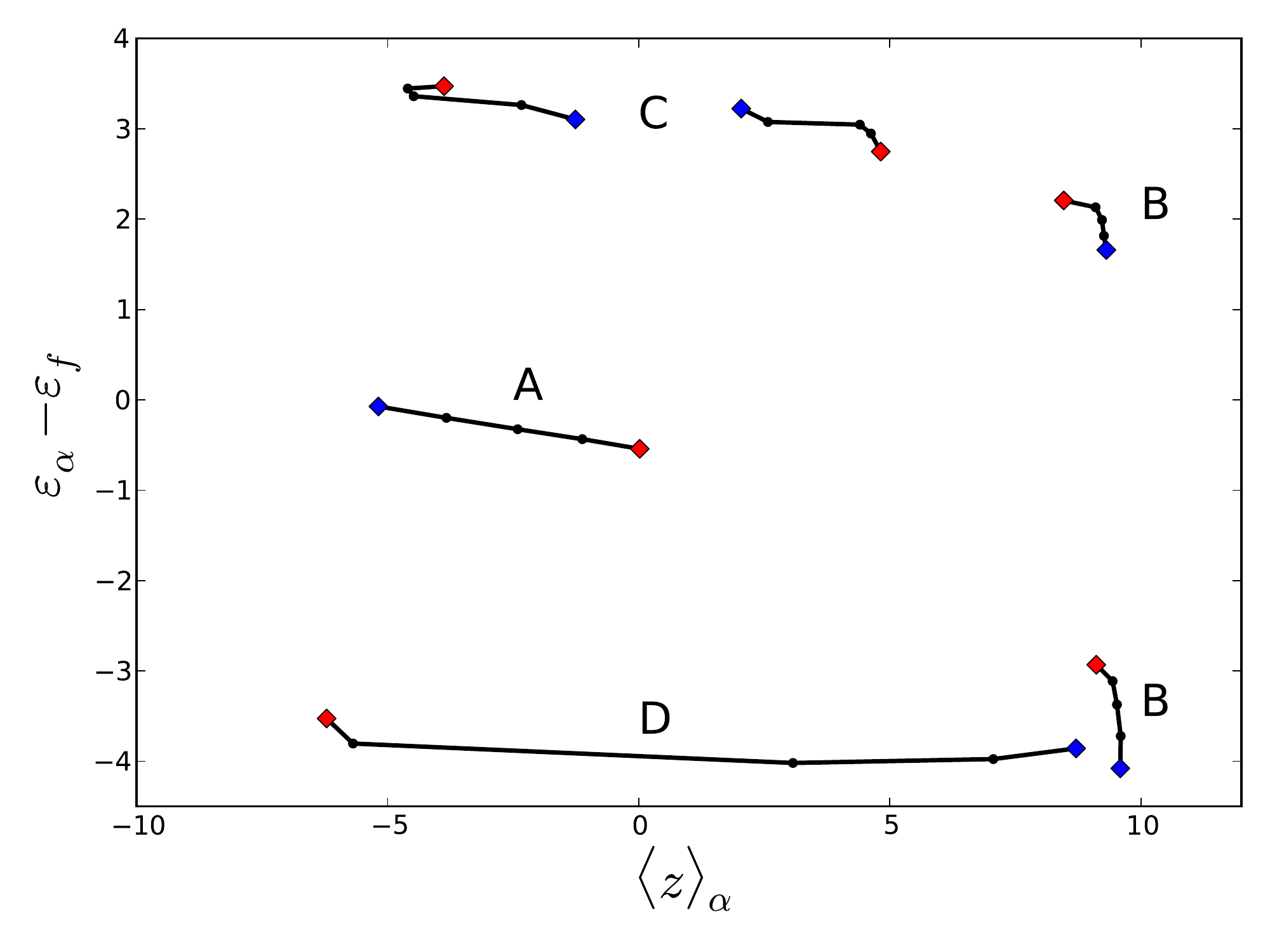} 
\caption{$K=1/2$ orbital energies and location parameters 
$\langle z \rangle$ as the Glider neck parameter changes.
The upper and lower panels show proton and neutron orbitals
respectively, for $|\varepsilon
-\varepsilon_f| < 4$ MeV.
Red diamonds:  Initial configuration \gpre; Blue diamonds:
\gpost.  See text for comments on the marked orbitals.
\label{Evsz} 
}
\end{center} 
\end{figure}

It is important for the diabatic treatment of the energy that the
evolution of the orbitals can be tracked across the scission shape changes.
This is hardly possible with the HFB wave functions constrained only by
shape.  With the HF wave functions and the two shape constraints  $Q_2$
and $N_{\rm neck}$, the orbitals evolve smoothly and one can identify the
individual orbitals in the two end-point configurations with little
ambiguity.  This is illustrated in Fig. \ref{Evsz}, 
showing how the orbital energies and their location vary as the 
neck size of the configuration decreases.  For most of the
orbitals (marked with `A'), the absolute value of $\langle z \rangle$ increases as the neck becomes
smaller.  This is exactly what we expect:  the bridge states
straddling both proto-fragments but become concentrated 
on one fragment or the other when the neck disappears. Orbitals
that are already localized on one of the fragments ( marked with
`B') hardly move at all.  Interestingly, there are two orbitals that
don't fit into the pattern.  The orbitals near the `C' marker move
in the opposite direction.  Undoubtedly, the reason is that there
are two orbitals at nearly the same energy that mix together.  When
diagonal energies become degenerate, the mixing becomes maximal.  This
seems to occur in a configuration close to $|168p\rangle$.  The
orbital evolution marked with `D' does not have any physical explanation.
Perhaps the chain of orbitals was incorrectly assigned.  

\subsection{Excitation energy}
In principle there are three contributions to the 
energy of the pre-scission configurations.   The first is
the ZQP energy depending only on the shape parameters and the $K$-partition.
The second contribution is
the excitation energy associated with 
quasiparticle excitations.
In this work we assume that it
can be computed from the difference in the single-particle energies
$\varepsilon_\alpha$  
of the ZQP configuration.
The last contribution, the collective kinetic energy,
will not be treated explicitly in our work here.

We relate the excitation energy of the configuration to its quasiparticle
spectrum using the standard grand canonical ensemble 
for occupation numbers $n_\alpha$,
\be
n_\alpha = \frac{1}{1 + \exp\left((\varepsilon_\alpha - \mu)/T\right)}.
\ee
Here $\mu$ is the chemical potential and $T$ plays a role like
temperature.  The qualification ``plays a role" is needed because a true
temperature is a property of a fully equilibrated system rather than a
system constrained by a specific $K$-partition.  The $T$ in the above
equation is only used to relate the partial occupation probabilities 
to the total excitation energy.  It is also important that
$\mu$ be adjusted to give the correct average particle number
in the ensemble.

We will calculate total excitation energies 
with using single-particle energies from both 
pre-scission  and post-scission
configurations $|166\rangle$ and $|168p\rangle$. We separate
the orbitals into two sets, $H$ and $L$ depending on the orbital's
location as indicated by $\langle z \rangle_\alpha$.  For the pre-scission
configuration, the orbital occupation numbers and single-particle
energies are denoted by a superscript ($^-{}$) as $n^-_\alpha$ and 
$\varepsilon^-_\alpha$, and similarly with ($^+$) for the post-scission
orbitals.   The  excitation energy of the pre-scission
configuration is given by  Eq. (\ref{Estar-eq}) with $n_\alpha =
n_\alpha^-$.  We make a preliminary division between the two nascent
fragments from the energies
\be
E^{*-} _{S} = \sum_{\alpha \in S} \left(n^-_{\alpha}- n^{0-}_{\alpha} \right)
\varepsilon^-_{\alpha}
\label{E_Lb}
\ee
where $S$ is the set $H$ or $L$.  
This changes to
\be
E^{*d}_{S} = \sum_{\alpha \in S} \left(n^-_{\alpha}- n^{0+}_{\alpha} \right)
\varepsilon^+_{\alpha}
\label{E_Ld}
\ee
after scission.
Here  the orbital occupation numbers are taken from pre-scission
occupation factors but the quasiparticle energies from the post-scission
configuration.  As mentioned earlier,
Eq. (\ref{E_Ld}) requires tracking 
individual orbitals along the scission path.  We saw in the last section
that this can be carried out fairly confidently, at least for
the orbitals near the Fermi energy.  More details of how we link the
pre-scission and
post-scission orbitals are given in the Appendix.
For Glider orbitals, our procedure satisfies
the check  $n^{0+}_\alpha = n^{0-}_\alpha$ required for a ZQP diabatic
path.

Table \ref{Estar-table} shows the energy partition calculated
by the above equations
for \gpre~and \gpost~ at initial excitation 
energies of 10 and 20 MeV.  For each set of orbital energies,
the parameter $T$ in Eq. (\ref{Estar-eq}) is chosen to reproduce
a given total $E^*$.  
The orbitals 
are assigned to $H$ or $L$ sets according to the sign of $\langle z
\rangle_\alpha$.
The average number of quasiparticles is given as $N_{qp}$ in the
Table.  
The final column in the table is the fraction of 
excitation energy in the heavy fragment, 
\be
f_H = \frac{E^*_H}{E^*}.
\ee

The first point to notice is that the excitation energy hardly
changes
during the diabatic evolution.
This is contrary to the expectation that diabatic transformations 
increase the internal energies of the system.  Evidently this
is not the case for Glider, which in fact evolves almost adiabatically
within its $K$-partition constraint.  However, we will see below an 
example of a diabatic evolution that explicitly injects excitation
energy into the final state fragments.

Another important observable is the energy sharing between
the post-scission fragments.  
We see from the Table that the quasiparticle excitation energy sharing 
is about the same
in the pre-scission 
(Ensemble I) and post-scission (Ensemble II) states.  To make 
contact with the scission-point statistical model, we also show
energy sharing in an ensemble based entirely on the final state
quasiparticle energy (Ensemble III).  Here the occupation
probabilities are computed with Eq. (\ref{Estar-eq}) without any
$K$-dependent chemical potential.  This ensemble shows some slight favoring
of the lighter fragment.   The computed energy fraction $f_H$ is shown in the 
bottom entries to the Table. Experimentally, there is a strong favoring
of the lighter fragment at this mass splitting.
It is attributed to the proximity of
the magic numbers 50 and 132 in the charge and mass numbers of the
heavy fragment.  Of course, our results should not be compared directly
with experiment because the ZQP energy has not been included.
\begin{table}[htb]  
\begin{center}  
\begin{tabular}{|c|cc|cccc|}  
\hline  
Ensemble & $T$  & $N_{qp}$ & $E^*$   &  $E^*_H$  & 
$E^*_L$  & $f_H$ \\
\hline
I & 0.893 &5.4 & 10.0 & 4.8 & 5.2 & 0.48 \\
& 1.164 & 8.7 & 20.0 & 10.4& 9.6 & 0.52 \\
II & 0.896 & 5.4 & 10.1 & 5.4& 4.6 &0.54  \\
  & 1.164& 8.7 &20.6 & 11.8  & 8.8  &0.57  \\
III & 0.923& 7.5 & 10.0 &4.6& 5.4 & 0.46 \\
  & 1.21& 10.7 &20.0 & 9.8 & 10.2 & 0.49 \\
\hline  
\end{tabular}  
\caption{Thermal energy associated with Glider at the scission point.
Ensemble I:  Eq. (\ref{Estar-eq}) with $n_\alpha^-, \eps_\alpha^-$ and
preserving $K$-partition on average; Ensemble II: Eq. (\ref{E_Ld}), diabatic
with occupation numbers from I; Ensemble III:  Eq. (\ref{Estar-eq}) 
with post-scission $n_\alpha^+, \eps_\alpha^+$ 
and unrestricted by $K$-dependent chemical potentials.  
There are two rows for each method giving results for starting energies
of 10 and 20 MeV.  The third column is the number of quasiparticles
$N_{qp} = \sum_\alpha |n_\alpha - n_\alpha^0|$.
The last two entries are from a scission-point statistical model.
The parameter $T$ and excitation
energies $E^*$ are in units of MeV.
\label{Estar-table}  
}
\end{center}  
\end{table}   

\section{Other examples}
\label{other}
Glider is perhaps one of hundreds of configurations that can
carry the \u~nucleus past the scission point. 
Any general conclusions would require investigating a representative
sample of them.  Toward this end, we have found several other
$K$-partitions that transition from shapes with distinct necks to
separated fragments.  Also, we present calculations with a different
energy functional to get some indication of 
which qualitative features of the dynamics are generic 
or strongly functional-dependent.

\subsection{Other $K$-partitions}
\begin{figure}[tb] 
\begin{center} 
\includegraphics[width=1.0\columnwidth]{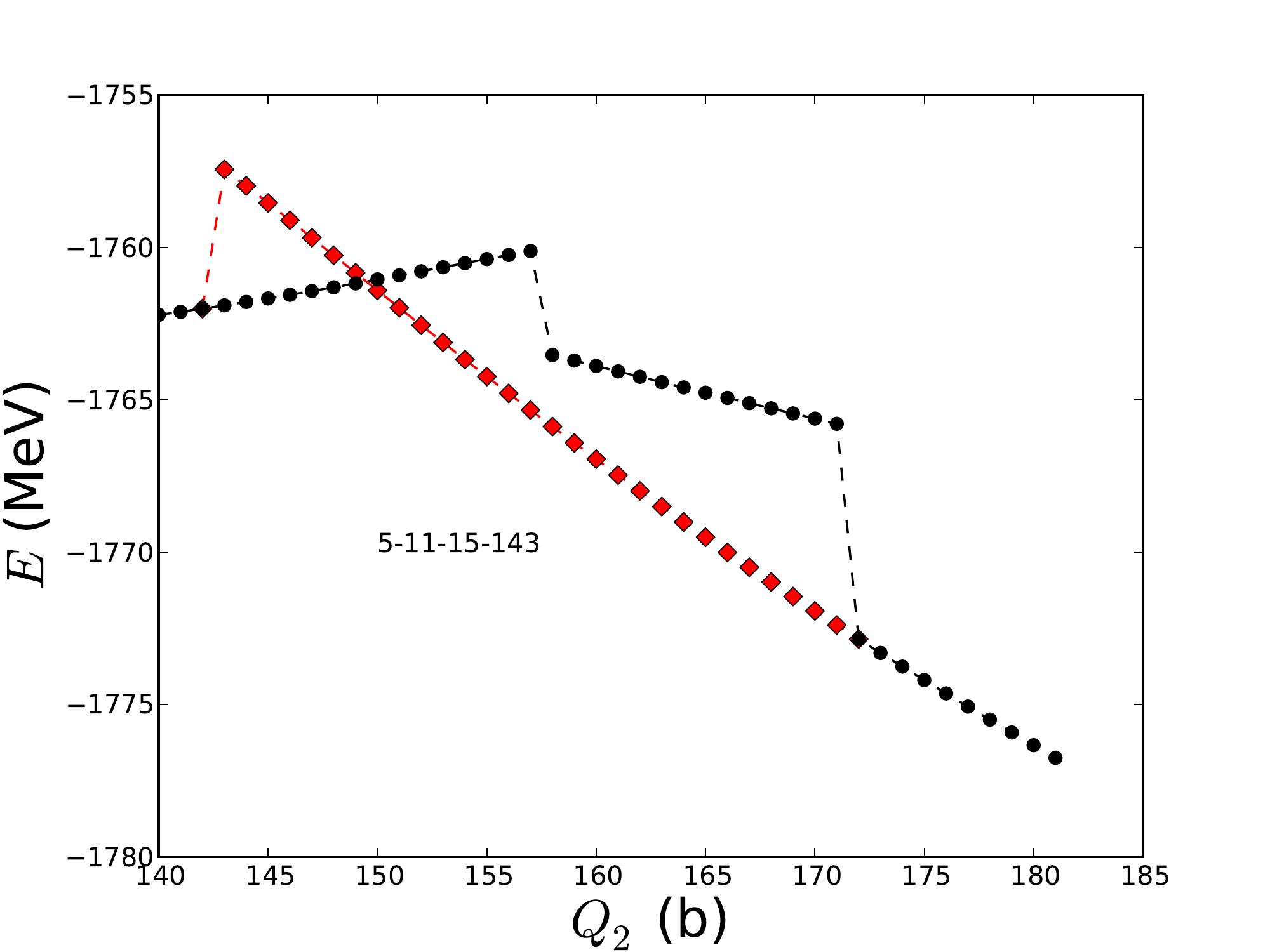} 
\caption{
Energy of the A configuration with the D1S energy functional and constrained by the
quadrupole field $Q_2$.  Black circles:  iteration from the left;
Red diamonds:  iteration from the right.
\label{QE-A}
}
\end{center}
\end{figure}

We consider here the three additional $K$-partitions labeled A,B,C  
in Table \ref{NK}.  We have two measures of the dissimilarity of
the configurations on either side of the scission.  The first measure
is the number of jumps in the $Q_2$-constrained path through the 
scission region.  In the case
of Glider, there is a single jump down stepping from the left and a
corresponding jump up stepping from the right; the two are very close in 
the $Q_2$ coordinate.
These numbers are compared with the other configurations in Table
\ref{zeta-t}.  
\begin{table}[htb]  
\begin{center}  
\begin{tabular}{|cc|c|cc|}  
\hline  
$K$-partition   &  Functional   & $N_j$ & $Q_2$ & $\zeta$\\
\hline  
Glider  & D1S & 1  & 152-162 & 2.2\\
A     & D1S & 2 & 149   &  4.1 \\
B     & D1S & 0 & 136   &  1.7 \\
C     & D1S & 0 & 134   &  2.4 \\
\hline  
Glider  & BCPM & 2  & 156-175 & 2.6\\
A     &  & 1 & 158   &  3.1 \\
B     &  & 1 & 141   &  1.7 \\
C     &  & 1 & 138   &  2.7 \\
\hline  
\end{tabular}  
\caption{Characteristics of the GCM paths through the scission point.
$N_j$ is the number of jumps in $N_{\rm neck}$  going from pre-scission
to post-scission shapes in steps of $\Delta Q_2 = 2$ b.
\label{zeta-t}  
}
\end{center}  
\end{table}   
The number of downward jumps range from zero for $K$-partition
B to two for A.  Partition A is an especially difficult case for 
constructing a path through the scission point.   Its
energy as a function of the single $Q_2$ constraint is shown
in Fig. \ref{QE-A}.
There are two configuration jumps stepping from smaller to larger $Q_2$,
and one jump stepping from the other side, The coexistence of two
configurations at the same $Q_2$ extends over a much larger range
($143 < Q_2 <  171$ b) than for the other cases.  Add the neck
constraint brings the two paths close together but there remains
a small region with two local minima.  

\begin{figure}[tb] 
\begin{center} 
\includegraphics[width=1.0\columnwidth]{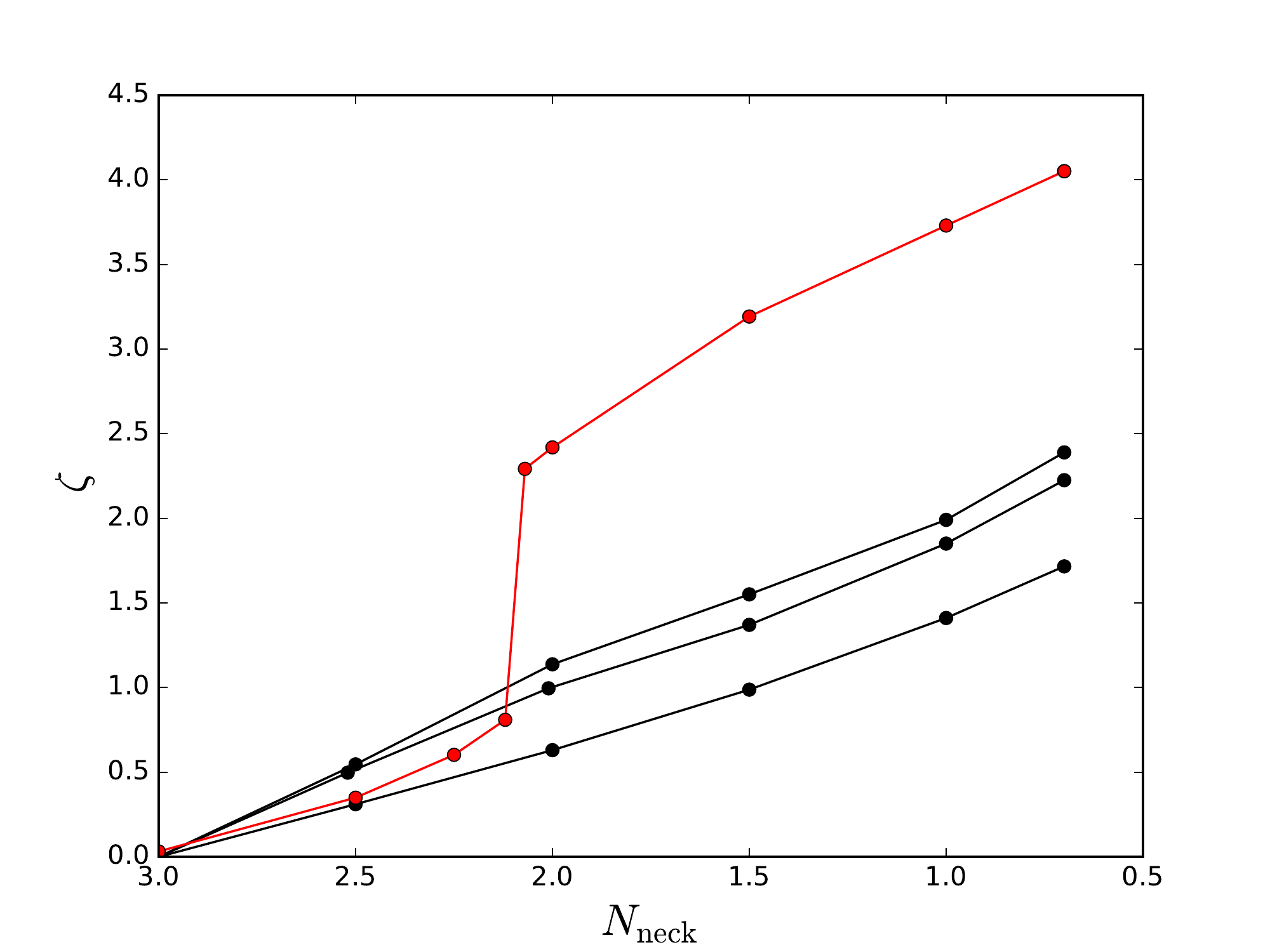} 
\caption{Overlap distance $\zeta$ from $N_{\rm neck} = 3.0$ to
0.7 for the four configurations Glider,A,B, and C 
with the D1S energy functional.  The A configuration is 
distinguished by red lines and markers.
\label{zeta-D1S}
}
\end{center}
\end{figure}

Another measure of difficulty in traversing the scission point is
the overlap distance $\zeta$.  Table \ref{zeta-t} shows $\zeta$
for a path through the scission point defined by $Q_2$ and the
neck size, with 
$N_{\rm neck}$ decreasing from 3.0 to 0.7.  For most cases the
distance is insensitive to the choice of $Q_2$.  However, varying
the $Q_2$ constraint along the path may make it somewhat shorter.
To get a sense of the smoothness of the diabatic
paths through the scission point we show $\zeta$ in Fig. \ref{zeta-D1S} 
as a function of the neck constraint.
One can see that partitions B and C are
quite similar to Glider, but A has a major rearrangement at
$(Q_2,N_{\rm neck}) \approx (149\,\,{\rm b},2.1)$.  We have traced
this behavior to the creation of a particle-hole excitation at
this point in the path.  A more detailed description is given in
the Appendix.

\subsection{The BCPM functional}
\begin{figure}[tb] 
\begin{center} 
\includegraphics[width=1.0\columnwidth]{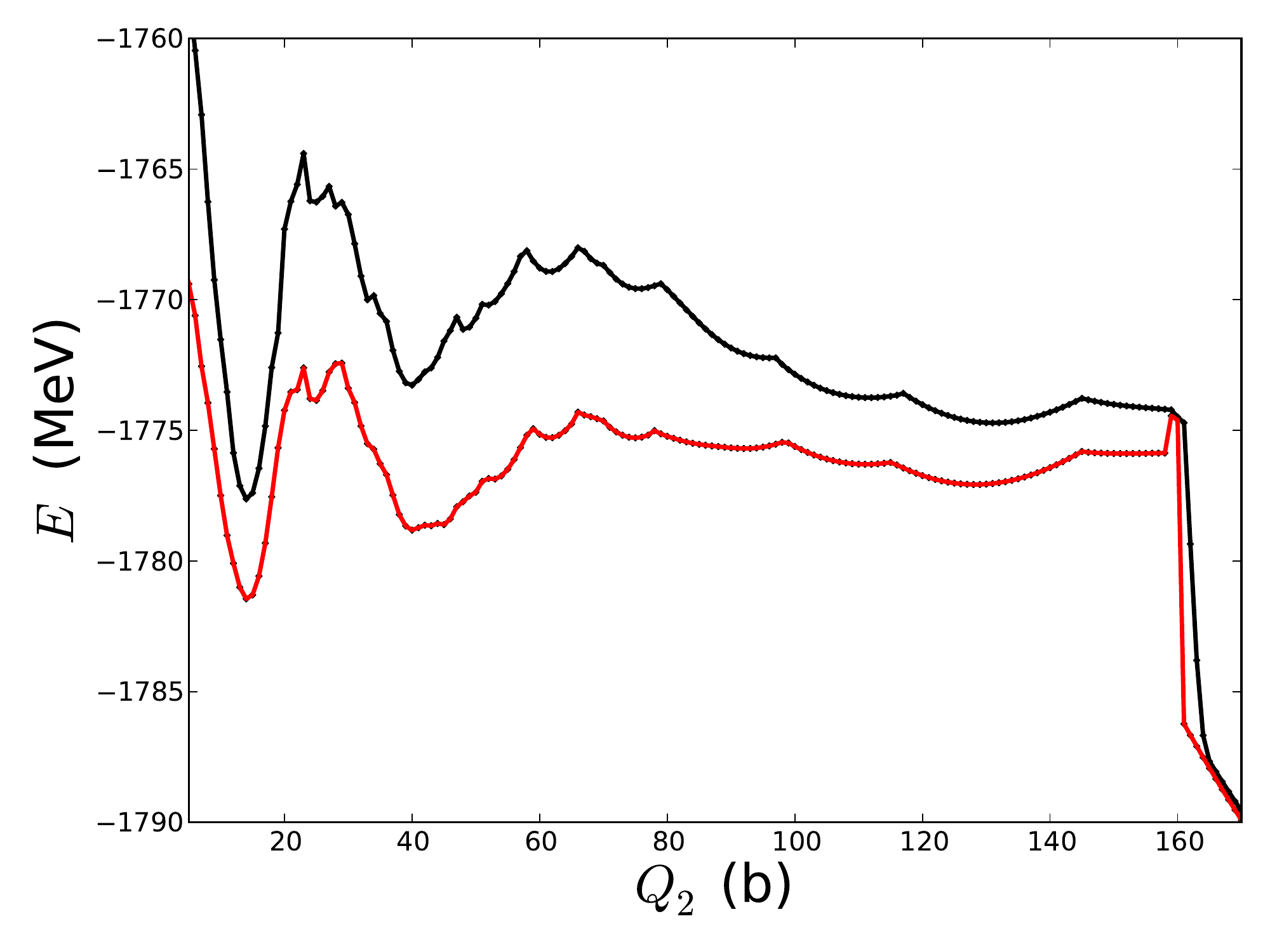} 
\caption{
Hartree-Fock Potential energy surfaces for \u~along the fission valley constrained by
$Q_2$. The energies functionals are the  D1S (black lines) and the
BCPM (red lines).
\label{bcpm-path}
}
\end{center}
\end{figure}

\begin{figure}[tb] 
\begin{center} 
\includegraphics[width=1.0\columnwidth]{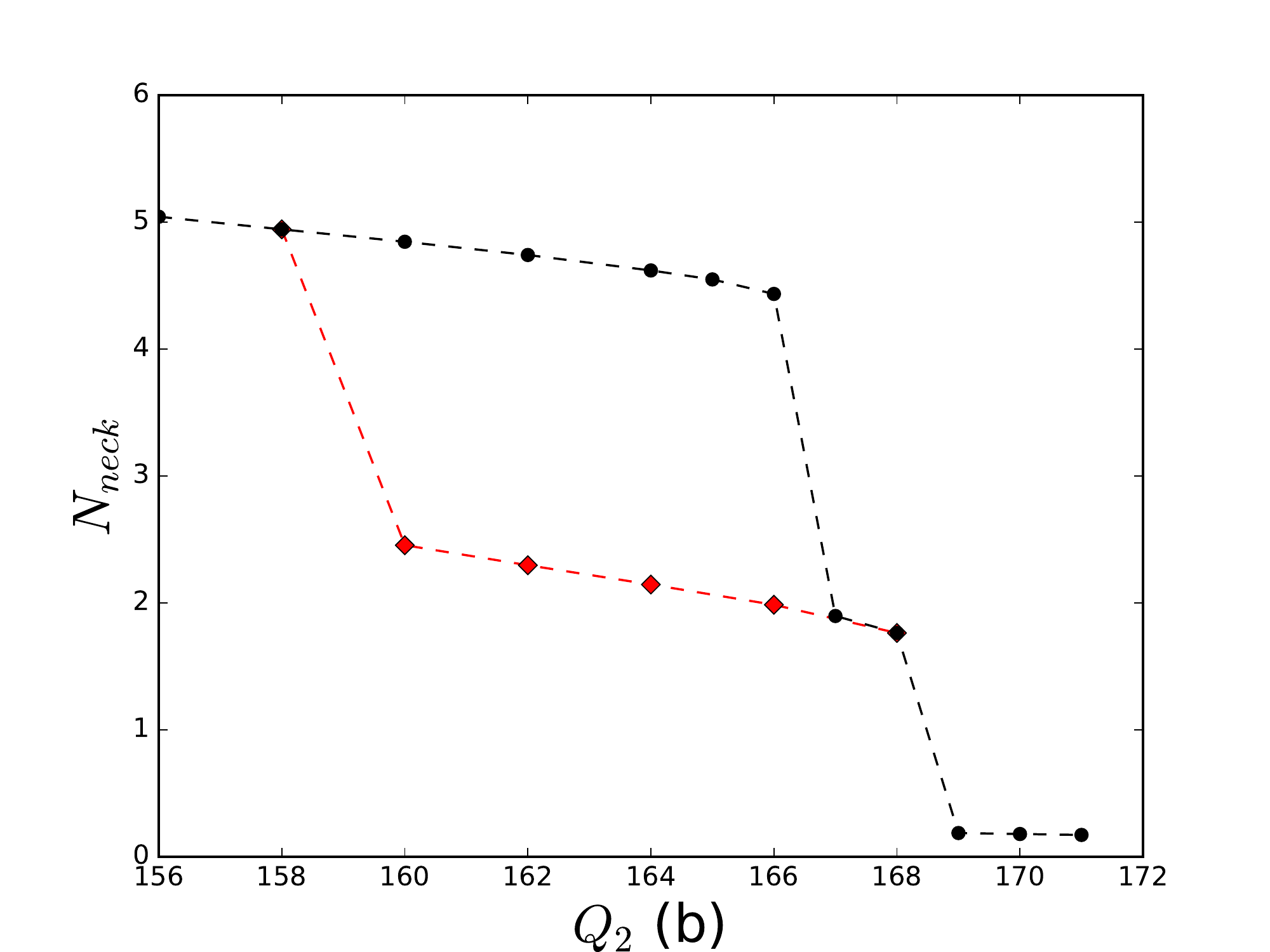} 
\caption{
\label{Qz} 
Glider neck size with the BCPM energy functional and constrained by the
quadrupole field $Q_2$.  As in Fig. \ref{QE-A},  iterations from the left and 
right are shown as black circles and red diamonds, respectively.
\label{n_neck2}
}
\end{center}
\end{figure}

\begin{figure}[tb] 
\begin{center} 
\includegraphics[width=1.0\columnwidth]{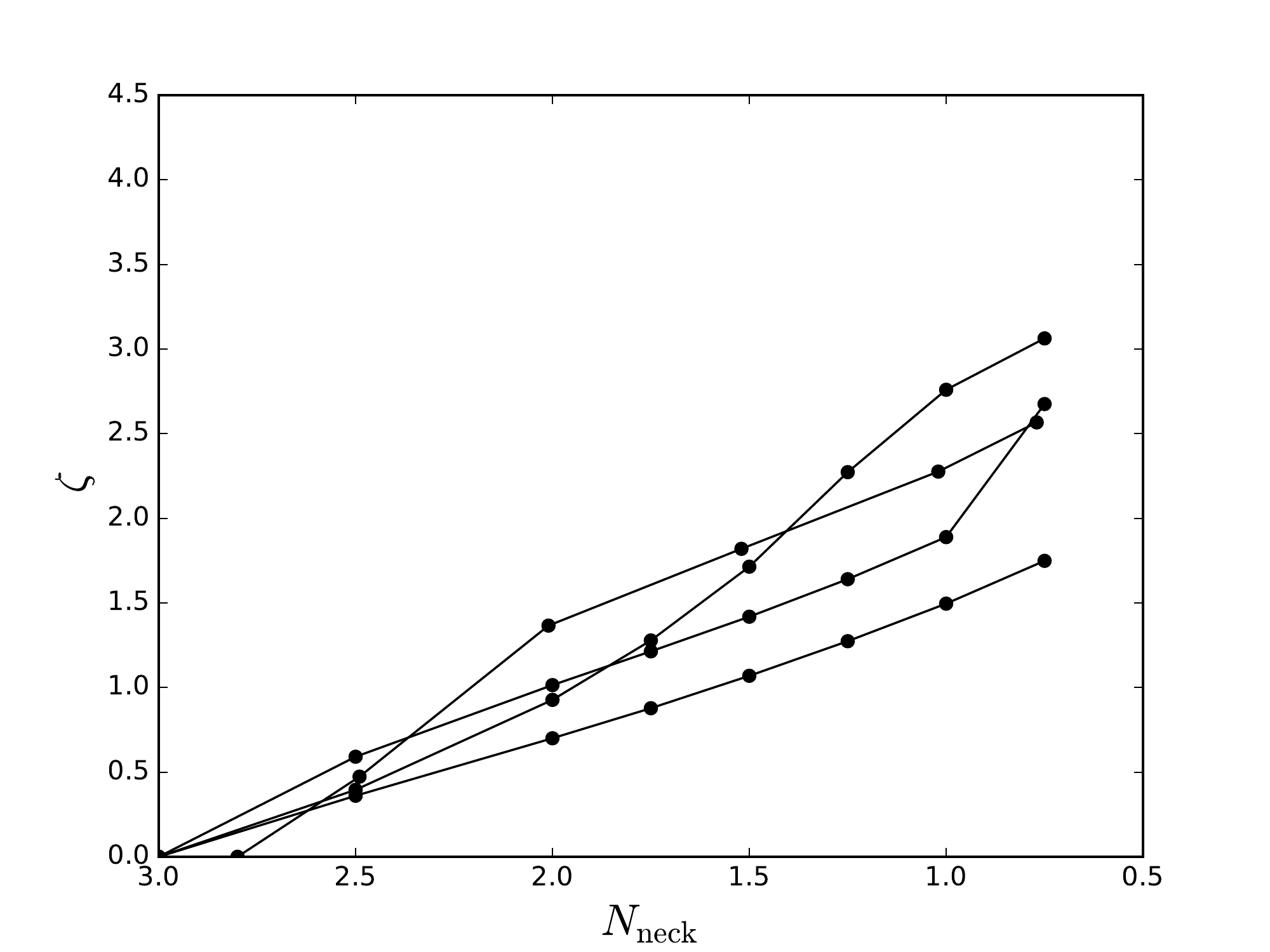} 
\caption{Overlap distance $\zeta$ from $N_{\rm neck} = 3.0$ to
0.7 for the four configurations Glider,A,B, and C 
calculated with the BCPM energy functional.  
\label{zeta-bcpm}
}
\end{center}
\end{figure}

We now carry out the same path analysis with the BCPM functional
\cite{ba13} which has been used in a previous fission study
\cite{gi13}.
The single-particle potential in BCPM is purely
local, giving a more realistic single-particle energy spectrum than the 
D1S. Despite the differences between the D1S and
the BCPM, the fission paths along the valley bottom are very similar
(Fig. \ref{bcpm-path}).
In particular, the sequence of $K$-partitions is identical 
starting from the ground state at $Q_2 = 14$ b to $Q_2 = 150$ b.  
However, closer to the scission point the paths are
far from identical.
Fig.
\ref{n_neck2}  shows Glider's \neck~vs. $Q_2$ calculated as in
Fig. \ref{n_neck}, but using the BCPM functional.  One sees that the
region of ambiguity is larger and that there is a sudden jump from
$N_{\rm neck} = 1.8$ to 0.2 that is not present in the D1S trajectory.  
Numerical data about the paths for all the partitions treated in the 
last section are tabulated in the bottom rows of Table V.  The $N_j$ for BCPM are all different from
the D1S values.  Nevertheless the overlap distances 
through the scission point are within 0.4 units of each other,
with the exception of partition A. Unlike the experience with D1S,
the BCPM permits one to construct a path through the scission point
with only two shape constraints.  Fig. \ref{zeta-bcpm} shows $\zeta$ as function
of \neck~as in Fig. \ref{zeta-D1S}.  The curves look quite similar to the better-behaved
D1S curves.   

\subsection{Excitation energies}
As a final task to explore the sensitivities to input assumptions we repeat the calculation of excitation energy production
and sharing during the scission.
The results for all configurations and both energy functionals 
are shown in Table \ref{Estar-t2}.
The pre-scission
and post-scission configurations are chosen at neck sizes
$N_{\rm neck} = 3.0$
and 0.7, respectively. The first thing to note is that the
effective temperature is higher for the D1S functional than for
the BCPM. This is not unexpected. Roughly speaking, the temperature
associated with a given excitation energy depends quadratically
on the effective mass in the energy functional.  The BCPM has an
effective mass of $m^*_{\rm bcpm} = 1$ while the D1S has an effective 
mass of about $m^*_{\rm D1S} = 0.67$.
The average temperatures in the above
Tables are 0.825 MeV and 0.728 MeV for BCPM and the square
of their ratio is 0.7, rather close to $m^*_{\rm D1S}/m^*_{\rm BCPM}$.
\begin{table}[htb]  
\begin{center}  
\begin{tabular}{|cc|cccc|}  
\hline  
Model &$K$-partition & $T$  & $f^-_H$ &   $f^d_H$  & $E^{*d}$\\ 
\hline
D1S &Glider& 0.9 & 0.41 & 0.57 & 10.6 \\
&A & 0.86& 0.31& 0.35& 18.9 \\
&B & 0.89  & 0.50  & 0.47 & 9.7\\
&C & 0.85 & 0.50 & 0.39 & 8.6\\ 
\hline
BCPM &Glider& 0.75 & 0.52 & 0.53 & 10.2 \\
& A & 0.695 & 0.29 & 0.39 & 11.9 \\
& B & 0.762  & 0.50  & 0.55 & 8.9\\
&C & 0.705 & 0.54 & 0.50 & 4.9\\ 
\hline
\end{tabular}  
\caption{Fraction of excitation energy in the heavy fragment 
and total final state excitation energy for the diabatic evolution 
Details of the path constraints are given in Table V. 
The parameter $T$ and excitation
energies $E^*$ are in units of MeV.  
\label{Estar-t2}  
}
\end{center}  
\end{table}   

The excitation energies shown in the table and how they are shared
between the fragments gives some idea of how much the scission-point 
dynamics affects these important quantities. Concerning
the total excitation energy, the final diabatic excitation energy can be
larger or smaller than the initial 10 MeV.    
 Excluding partition A under D1S, the average
change is -8\%, largely driven by the 50\% decrease in partition
C under BCPM.  The corresponding standard deviation of the energies
is $\pm 2$ MeV.  The other question is how the excitation energy get shared between
the two fragments.  For the 7 normal cases in Tables VI, the
average fraction to the heavy fragment $f_H$ and its standard 
deviation is $0.48\pm0.07$.  The experimental $f_H$ depends strongly
on the mass splitting, and it can be reproduced in at least one
version of the scission-point statistical model \cite{al18}.  There
they find $f_H\approx 0.36$ at $E^*=10$ MeV; in our Table VI  3 out of the 8 partitions have energy fractions close to that
value.

\section{Conclusion and outlook}  

At the beginning of this study, we hoped that the constraint
on $K$-partitions would be powerful enough to permit
construction of paths through the scission point using 
a single shape constraint.  This condition is met in
two of the configurations, namely the ones that have no jumps
in Table \ref{zeta-t}.  But typically two constraints are needed and even that
is inadequate for partition A under the D1S energy functional.

This makes it  much harder to build a basis
of configurations that could be treated with  standard 
many-body techniques.
Still, it is reasonable to assume the scission paths with
short lengths (as measured by the overlap distance) will dominate
in the decays.  The more lengthy paths might be ignored in making
first estimates of decay rates.  It is intriguing to note that in
the time-dependent approach to fission dynamics it has been found
that there can be an important bottleneck at the scission point \cite{bu15}.
There it was found that the nucleus evolved smoothly to the scission
point but remained there for a variable amount of time ranging up
to $\tau_{\rm scission} \approx 1.4\times 10^4$ fm/c.  This implies
that the pre-scission configurations might have widths as small as
$\Gamma \approx \hbar /\tau_{\rm scission} \approx 10$ keV.
Decay widths of the order 10 keV and higher would be consistent
with the measured autocorrelation function for the $n+^{235}$U 
fission cross section, which does not show any systematic 
correlation on lower energy scales that could be attributed
to the scission decay width \cite{be18b}.

As a next step in the present program, we would like to estimate
Hamiltonian matrix elements between configurations along the 
scission path.  In this respect, it will be quite helpful to
be guided by the $\zeta$ distances when setting up the GCM space of
intermediate configurations. It should then be possible to
calculate
decay widths of configurations near the scission point following
the GCM methodology outlined in Ref. \cite{be19}.  This could
be carried out with the present computer codes, but it would be
desirable to include collective momentum variables among the
generator coordinates.  Once one has the tools in place to 
calculate decay widths,  it is a simple extension to the calculation
of branching ratios by decay widths of different final states.
For example, it would permit a fully microscopic theory of the 
distribution of kinetic energy in the final state (TKE).

In this study we have also used the diabatic paths to 
estimate the transport and changes in excitation energy across
the scission point\footnote{See Ref. \cite{ta17} for another
microscopic treatment of excitation energy transport.}. This is very relevant to the scission-point
statistical model describing the distributions in mass yields 
and excitation energies of fission fragments.  In that model, it is
assumed that the scission itself has no dynamics role.  This
is belied by partition A which undergoes a 2-quasiparticle excitation
on its scission path.  However, this may be an anomalous case
and it seems that ZQP paths are much more likely.  For those paths,
the average decrease is not much compared to all of the
other uncertainties, and the variance is also small compared to other
sources of fluctuation in the excitation energy.  The conclusion is
that ignoring diabatic dynamics remains an accepting approximation
in justifying the scission-point statistical model.
However, no firm conclusions can be drawn until a more representative
sample of partitions can be examined.

\section*{Acknowledgment}
We thank J. Randrup for providing data from the calculations in Ref.
\cite{al18}. 
This work was performed under the auspices of the U.S. Department of Energy
by Lawrence Livermore National Security, LLC, Lawrence Livermore National
Laboratory under Contract DE-AC52-07NA27344. Funding for travel was provided
by the U.S. Department of Energy, Office of Science, Office of Nuclear
Physics under Contract No. DE-AC02-05CH11231 (LBNL), through the University
of California, Berkeley.  The work of LMR was partly supported by Spanish
MINECO grant Nos. FPA2015-65929 and FIS2015-63770.

\section{Appendix}

The possibility of creating particle-hole excitations
along the fission path, and especially near scission, is of considerable
interest in fission theory \cite{Bernard2011}. The excitation
of intrinsic states in the latter stages of fission has a direct bearing
on the excitation energy imparted to the fragments at scission, and
the energy available for neutron emission. In this appendix, we discuss
in more detail how we track the orbitals along the fission path by 
their overlaps, finding a particle-hole excitations on the
diabatic path.  

 In a HF formulation of the problem, the orbitals
are defined by the matrix $W^q$ diagonalizing the single-particle
mean-field Hamiltonian $H^q$. Here $q$ are the constraints imposed
in the minimization of the energy functional.  The overlaps of
orbitals $\alpha,\alpha'$ at different $q$   is simply the dot product 
of the two vectors
\be
\langle \alpha q | \alpha' q'\rangle = \sum_i W^{q*}_{i,\alpha}
W^{q'}_{i,\alpha'}.
\ee
As noted in section II, the calculations in this paper are carried
out within the HFB framework, with the HF wave functions produced by
constraining the pairing condensate to be small.  In that case,
the columns of the $W$ matrix map into rows of the $U$ matrix if
the orbital is unoccupied and into rows of the $V$ matrix if the orbital
is occupied.
Both cases are covered 
by the formula
\begin{equation}
\begin{aligned}
\langle \alpha q | \alpha' q'\rangle  =
\sum_{k}\left[U_{\alpha k}^{q*}
U_{{\alpha'}k}^{q^{\prime}}+V_{\alpha k}^{q*}V_{{\alpha'}k}^{q^{\prime}}\right]\end{aligned}
\label{eq:UUplVV}
\end{equation}
when the occupation numbers are the same.
If the occupation number changes, the nonzero amplitudes reside in
the $U$ matrix for one of the orbitals and in the $V$ matrix for the other.
Then the overlap may be computed as
\begin{equation}
\begin{aligned}
\langle \alpha q | \alpha' q'\rangle =
\sum_{k}
\left[U_{\alpha k}^{q*}V_{\alpha' k}^{q^{\prime}}
-V_{\alpha k}^{q*}U_{\alpha' k}^{q^{\prime}}\right]\end{aligned}.
\label{eq:UVmnVU}
\end{equation}

In practice, 
the larger in absolute
value of Eqs. (\ref{eq:UUplVV}) and (\ref{eq:UVmnVU}) is adopted
as the optimal overlap. For each value of the $K$ quantum number,
the optimal overlaps 
$|\langle \alpha q | \alpha' q'\rangle|$
are calculated for all possible orbitals $\alpha,\alpha'$. 
The
overlaps are then sorted from highest to lowest in absolute value.
Proceeding down the list of sorted overlaps, the orbitals 
for the
largest overlap 
are considered matched and taken out of consideration. The next largest
overlap which does not involve those orbitals already taken out of consideration
gives the next matched pair of levels. This process is repeated until
all orbitals have been matched.

\begin{figure}
\includegraphics[width=1.0\columnwidth]{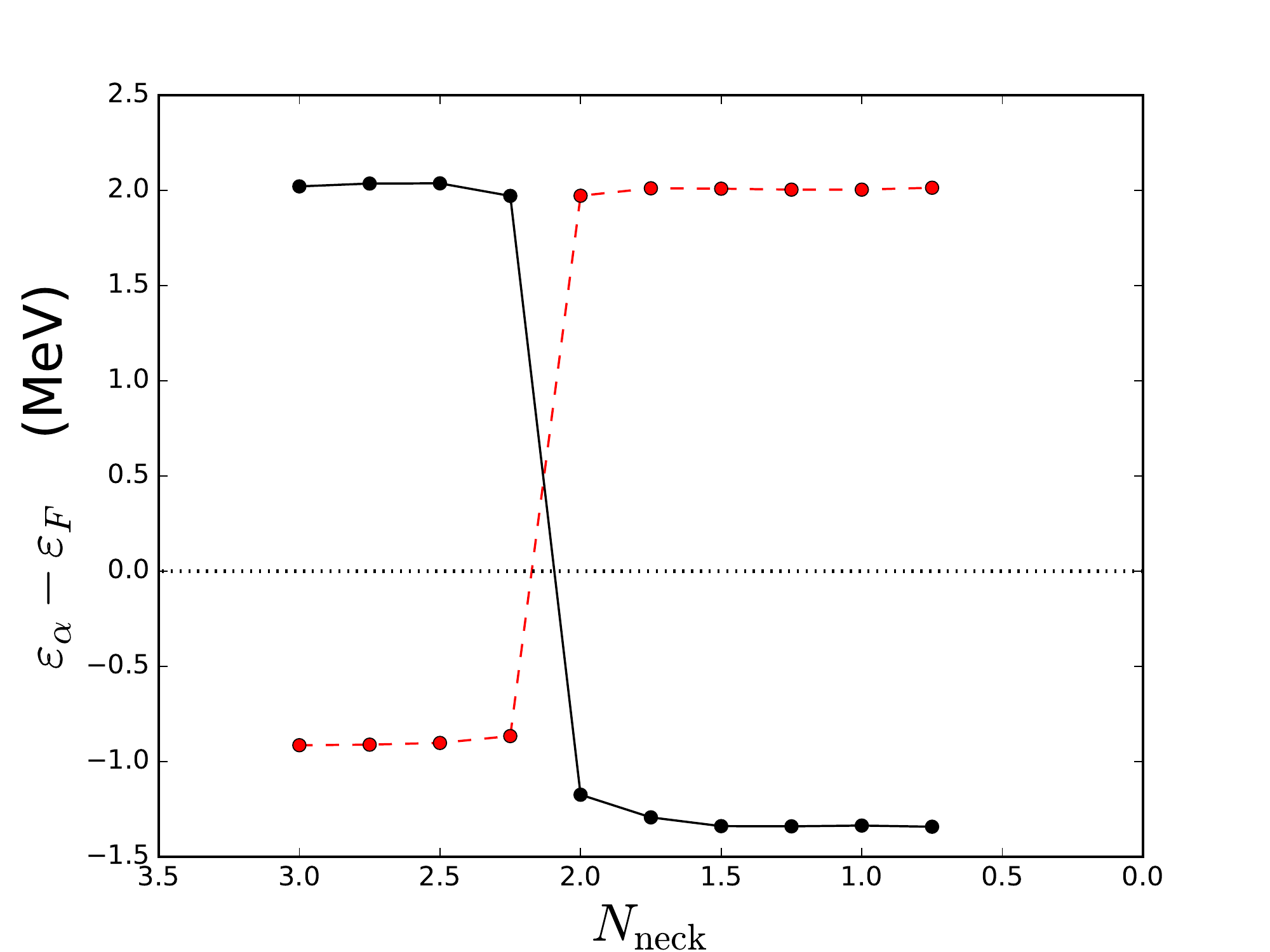}
\caption{\label{fig:5-11-15-143-149r-eps}Single-particle energies, relative
to the Fermi energy, plotted as a function of neck size for the two
lowest $K=5/2$ proton orbitals for the configuration $A$ at $Q_{2}=149\:\textrm{b}$.}
\end{figure}

We now examine in detail the case where we found a possible particle-hole
excitation, namely  configuration $A$ at $Q_{2}=149\:\textrm{b}$ in
table V, calculated with the D1S interaction. The neck size was constrained
from $N_{\textrm{neck}}=3$ down to 0.75 in steps of $\Delta N_{\textrm{neck}}=0.25$.
After calculating orbital overlaps according to the above procedure, the
first
two proton orbitals in the $K=5/2$ block were found to change occupations
when going from $N_{\textrm{neck}}=2.25$ to 2. The tracked single-particle
energies for those two orbitals are plotted as a function of neck size
in Fig. \ref{fig:5-11-15-143-149r-eps}. One orbital switches from occupied
to unoccupied as the neck shrinks, while the other switches from unoccupied
to occupied. This swap in occupation probabilities can be interpreted
as a particle-hole excitation between the two configurations.
Interestingly, both orbitals reside largely in the same fragment, but
one has orbital angular momentum $M$ nearly parallel to the spin angular
momentum, $K = M +1/2$ while the spin in the other is mostly antiparallel.
The spatial densities of the orbitals in the HF
representation are given by
\begin{equation}
\begin{aligned}\rho_{\alpha}\left(z\right) & =
\sum_{i,k}\delta_{K_i,K_k} \delta_{M_i,M_k}
W^*_{i,\alpha}W_{k,\alpha}
\end{aligned}
\ee
$$
\times\int dx dy\,\,
\phi^*_{i}\left(x,y,z\right)\phi_{k}\left(x,y,z\right)
\label{eq:spdens}
$$
where $\phi_k(\vec r)$ are the basis orbitals.  In the HFB representation
we replace $W^*_{i,\alpha}W_{k,\alpha}$ in the above equation
by $U^*_{\alpha,i}U_{\alpha,k}+V^*_{\alpha,i}V_{\alpha,k}$, similar to our
treatment of Eq. (11).

The densities of the
two $K=5/2$ proton orbitals are shown in Fig. \ref{fig:totdensity}.
The upper panel shows the densities in the orbital just below the Fermi level
at
$N_{\rm neck} = 2.10$ (dashed red line) and $N_{\rm neck} = 
2.09$ (solid black line).  The lower panel shows the corresponding
densities of the orbital just above the Fermi level.
It may be seen that the orbitals have very different spatial character:
one has a single lobe centered near the middle of the fragment,
while the other is extended over a much larger range of $z$ and
has three lobes. 
\begin{figure}
\includegraphics[width=1.0\columnwidth]{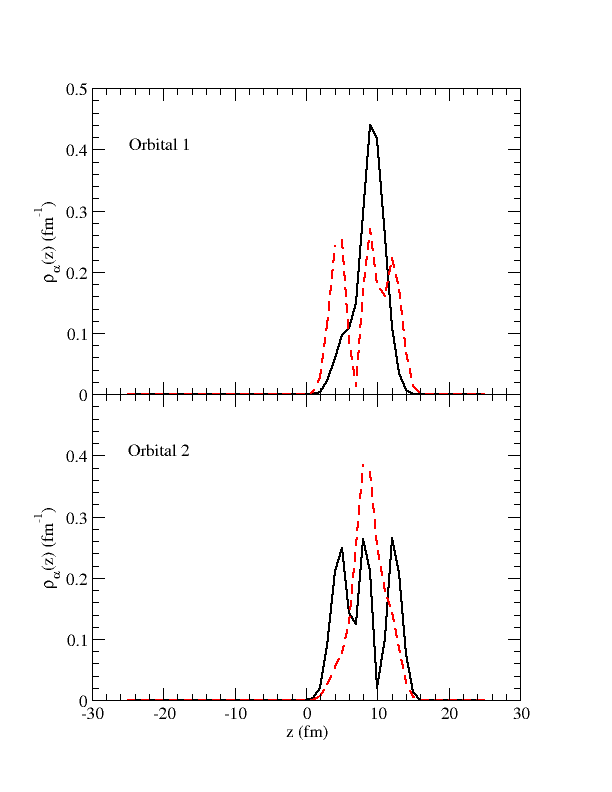}
\caption{\label{fig:totdensity}Single-particle densities for the 
two $K=5/2$ proton orbitals closest to the Fermi energy 
for the configuration $A$ at $Q_{2}=149\:\textrm{b}$  Dashed
red lines:  $N_{\rm neck} = 2.10$; solid black lines: 
$N_{\rm neck} = 2.09$.
}
\end{figure}

The tracking results described above has an implicit dependence on the
step size along the scission path.  The diabatic evolution is (somewhat
imprecisely) defined as the tracking with large step sizes as in
Fig. \ref{fig:5-11-15-143-149r-eps}.  For small step sizes, one would expect an adiabatic evolution.
That is, the orbitals will keep their position as ordered by their
single-particle energies.  The level crossings become avoided 
crossings, suppressing particle-hole excitations,  
except if prevented by symmetries.  Indeed, we found this to be the
case here.  The particle-hole transition takes place near
$N_{\rm neck} = 2.1$.  We examine the orbital matched between 
two configurations around that point and separated by some 
$\Delta N_{\rm neck}$.   
For $\Delta N_{\textrm{neck}}\lesssim0.07$,
the overlaps in Eq. (\ref{eq:UUplVV}) are favored leading to a smooth
transition across the critical neck size without p-h excitation. For
$\Delta N_{\textrm{neck}}\gtrsim0.07$, however, the overlaps in Eq.
(\ref{eq:UVmnVU}) are favored, resulting in a swap of occupation
probabilities and a corresponding p-h excitation. Although the calculations
in this paper are entirely static, the dependence on $\Delta N_{\rm neck}$
mimics the range of dynamic evolution at scission, where smaller $\Delta N_{\textrm{neck}}$
values can be identified with a slow process, and larger $\Delta N_{\textrm{neck}}$
values with a faster one.  

However, there one aspect of the Landau-Zener avoided-crossing picture in this
example that remains a puzzle.  Namely, the energies of the orbitals
should smoothly pass by each other as a function of the shape parameter, 
except for a small region of the avoiding crossing.  In fact, we
find that the energies remain nearly constant along the scission path.
Somehow, the wave functions are exchanged without the energies coming
close together.  Clearly, this aspect of the dynamics needs further
study.

\end{document}